# Quasi-Steady-State and Related Cosmological Models:
# A Historical Review[*]

Helge Kragh

Centre for Science Studies, Aarhus University, Denmark

**Abstract.** Since the emergence in the late 1960s of the standard hot big-bang theory, cosmology has been dominated by finite-age models. However, the rival view that the universe has existed for an indefinite time has continued to be defended by a minority of researchers. This view has roots far back in history and in the 1950s and 1960s several models were proposed in opposition to the big-bang paradigm. The most important of the alternative models, the steady-state cosmology proposed in 1948, was uniformly expanding rather than exhibiting a cyclical behaviour. In a much revised version it was developed into the quasi-steady-state cosmological model (QSSC) of the 1990s. From a historical point of view, this model, and a few other related models, can be seen as the latest examples of a tradition in cosmological thought that goes back to ancient Greece. The paper describes the background and development of the QSSC model.

## 1. Early conceptions of a cyclical-eternal world

The notion of an eternal universe, either in the steady-state version or some cyclical version, is as old as the idea of a created universe. Many of the ancient cosmo-mythologies imagined a future cosmic catastrophe represented by a huge battle between the good and evil forces of nature. The catastrophe did not necessarily imply an absolute end of the world, though, for out of the ashes of the old world a new might rise. In some cultures, notably in India, the process was thought to go on endlessly, with an eternal change between creative and destructive phases. This is the archetypical conception of the cyclical universe, an idea which has fascinated humans throughout history and can be found in mythical as well as scientific cosmologies right up to the present [1; 2].

---

[*] Slightly revised version of contribution to Rüdiger Vaas, ed., *Beyond the Big Bang: Prospects for an Eternal Universe*, a book to be published by Springer, but which has now been delayed for several years.

Cyclical conceptions were well known in ancient Greece, where models of this kind were postulated by pre-Socratic philosophers such as Amaximander and Empedocles (in the sixth and fifth centuries B.C., respectively). According to Empedocles, the cosmos was governed by two polar motive forces, metaphorically called "Love" and "Strife." The changes between dominance by Love and Strife proceeded eternally, corresponding to continual creations and destructions of the world. Only if there were a certain balance between the cosmic forces would life processes be possible [3]. Also the later Stoic philosophers adopted the idea of temporally multiple universes, which they associated with thermal phenomena. What Empedocles poetically had named Love and Strife was conceived more mechanically, as condensation and rarefaction. The world was seen a gigantic sphere oscillating through cycles of expansion and contraction in the void surrounding it. Chrysippus, a leader of the Stoic school in Athens in the third century B.C., is said to have believed that "after the conflagration of the cosmos everything will again come to be in numerical order, until every specific quality too will return to its original state, just as it was before and came to be in that cosmos" [4, p. 202].

Speculations of roughly the same type were considered by many later natural philosophers, including Isaac Newton and Immanuel Kant. With the emergence of thermodynamics in the mid-nineteenth century, the new science of heat was applied to the universe at large. While the second law was used to defend the notion of a finite-age, irreversibly evolving cosmos, other thinkers argued from the first law that the universe was eternal and possibly cyclic. The scenario of a "heat death" (*Wärmetod*) based on the law of entropy increase was vigorously resisted by philosophers of a materialistic or positivistic orientation who much favoured the alternative of eternal recurrence [5]. They typically argued that even though parts of the universe may be running down, there will always be other parts in which constructive counter-entropic processes dominate, the result being a cyclic, regenerating universe with no heat death. In effect, they claimed that the universe is a maximally large *perpetuum mobile*.

According to Oswald Köhler, an author and amateur astronomer, the world would approach a state of high entropy, yet it would only be a prelude to a new world. The universe, he wrote, was "like a clock that was never wound, that has run for all eternity and will run for all eternity." It was indeed "the true and unique *perpetuum mobile*" [6, p. 380]. Although parts of the cosmic machinery would continually break down here and there, at other places it would be repaired with the same speed, leaving the universe as a whole in an eternal state of equilibrium. Köhler and his kindred spirits (to whom belonged notables such as Friedrich Engels, Ernst Haeckel and Friedrich Nietzsche) believed that the universe was in a dynamic steady state and that this was a necessary consequence of the indestructability of matter and the eternity of time.

Cosmological ideas in roughly the same tradition were defended in the early decades of the twentieth century by a small group of physicists, chemists and astronomers who either ignored or resisted the new approach to cosmology based on the theory of general relativity. According to Svante Arrhenius, Emil Wiechert, Walther Nernst, Robert Millikan and William MacMillan, the universe was infinite in both space and time [7]. Strongly opposed to the gloomy prediction of the heat death, they suggested various mechanisms to maintain an eternal, ever-creative universe. Interestingly, Nernst's mechanism was based on the hypothesis of a zero-point cosmic radiation, the first idea of its kind [64]. When the expansion of the universe was established about 1930, Nernst and his allies similarly suggested alternative explanations of the galactic redshifts based on a static universe. However, many of their ideas were speculative and they failed to develop them into proper cosmological models. What mattered was the result, an eternal universe in equilibrium between organization and dissipation processes. This class of cosmological ideas was effectively based on what was called, in the later steady-state model, the perfect cosmological principle. Thus, according to the Chicago astronomer MacMillan, "the distribution of matter throughout space is uniform in the sense that […] this portion of the physical universe which comes under our observation is not essentially peculiar." Moreover, he denied "that the universe as a whole has ever been or ever will be essentially different from what it is today" [8, p. 99].

Following Albert Einstein's pioneering work of 1917, mainstream theoretical cosmology came to mean models based on the cosmological field equations of general relativity. Whereas Einstein's original model was static, the works of Alexander Friedmann and Georges Lemaître in the 1920s demonstrated the possibility of a wide class of dynamical models in which the space curvature changed in time. Friedmann was primarily interested in the mathematical aspects of dynamical relativistic models, not in the physics of the actual universe. Yet he seems to have been particularly fascinated by the possibility of a cyclical universe, for which he calculated a period of about $10^{10}$ years, based on the assumption of a zero cosmological constant ($\Lambda = 0$) and the mass of the universe being $M = 5 \times 10^{21}$ sun masses. In a book of 1923 he wrote about such a model that "The universe contracts into a point (into nothing) and then increases its radius from the point up to a certain value, then again diminishes its radius of curvature, transforms itself into a point, etc." He added that "This brings to mind what Hindu mythology has to say about cycles of existence" [9, p. 109]. Although the relativistic version of the cyclical universe fitted nicely into the age-old tradition of recurring worlds, from a conceptual point of view it was a radical novelty: the universe itself evolved in vast cycles, interrupted by cosmic singularities where the contraction into nothing ($R = 0$) was followed by expansion into a new world.

Models of this kind were occasionally considered in the 1930s, after Hubble's discovery of the redshift-distance relation had given observational support to the notion of the expanding universe [10, pp. 401-03; 65]. Lemaître considered cyclic solutions in a paper of 1933, but only to conclude that they were ruled out observationally. Yet he seems to have shared Friedmann's fascination with such models, which "have an indisputable poetic charm and make one think of the phoenix of the legend" [11, p. 84]. There were no good reasons to prefer an ever-oscillating universe to other models, and there were cosmologists who positively disliked the idea. Eddington admitted that he was "no Phoenix worshipper" and thought that it was "rather stupid to keep doing the same thing over and over again" [12, p. 86].

One major disadvantage of Friedmann-like cyclical models was that the universe – or temporally separated series of universes – passed through an infinity of singularities at which physics broke down. It was generally agreed that the cosmic singularity, corresponding to an infinite mass density, could not be part of the real world and must somehow be a mathematical artifact. If one singularity was a problem, an infinity of them was a scandal. In the spring of 1931 Einstein discussed a closed model of the Friedmann type with $\Lambda = 0$ in which the universe contracted to a point and subsequently expanded [13]. The model is sometimes referred to as the Einstein oscillatory model, but Einstein only considered a single cycle. He was convinced that the singularity was non-physical and believed that it would disappear if the standard assumption of homogeneity was abandoned.

Oscillatory, non-singular models were a preferable alternative, not only for conceptual reasons but also because they might evade the troublesome paradox of a universe that was younger than its constituent parts. Harold Spencer Jones, the later astronomer royal, thought of the possibility in 1934, and about the same time Willem de Sitter expressed his preference of a universe which "may have contracted during an infinite time and, after passing through a minimum a few thousand million years ago, started to expand again" [14, p. 414; 15, p. 708]. The problem with such non-singular models, as seen from the perspective of the 1930s, was that apparently they were not among the solutions to the Einstein cosmological field equations.

## 2.  The steady-state theory of the universe

In 1948 there appeared a strong alternative to the evolutionary cosmologies based on general relativity, in the form of the steady-state theory. The theory was proposed in two different versions, one by Fred Hoyle and the other by Hermann Bondi and Thomas Gold [16; 17]. Both versions adopted as a starting point the "perfect cosmological principle," meaning the assumption that as far as the large-scale features of the universe are concerned there is neither a privileged position nor a privileged time: the universe is spatially *and* temporally homogeneous. Since Hoyle, Bondi and Gold accepted the expansion of the universe, they were forced to introduce the radical hypothesis that matter is continually created throughout the universe, assumedly in the form of hydrogen atoms or protons and electrons (without the hypothesis the matter density would decrease over time, contrary to the perfect cosmological principle). As shown by Hoyle, it followed from the theory that the average density of matter was given by $\rho = 3H^2/8\pi G$, where $G$ is Newton's gravitational constant and the Hubble parameter $H$ was a true constant, contrary to the situation in the big-bang models (where $H$ is a measure of the age of the universe). To cancel the thinning out of matter as a result of the expansion, it was necessary to postulate creation of new matter at a rate of $3\rho H$ or about $10^{-43}$ g/cm$^3$ s. This exceedingly small value, corresponding to the formation of three new hydrogen atoms per cubic meter per million years, was far too small to have direct observational effects. Bondi and Gold further concluded that the metric of the steady-state universe must be of the same type as in de Sitter's model of 1917, i.e., a flat space (curvature parameter $k = 0$) expanding exponentially. It followed that the deceleration parameter, which is a dimensionless measure of the slowing down of the expansion and given by $q_0 = -\left[\ddot{R}/(RH^2)\right]_0$, had the value $q_0$ = -1.

Shortly after the publication of the papers of Hoyle, Bondi and Gold, the steady-state theory was met with strong opposition, the result being a protracted controversy that lasted until the mid-1960s [18]. The critics of the theory accused it of building on implausible hypotheses (such as the continual creation of matter) and being unable to account for observations. In order to counter these and other objections to the steady-state theory, Hoyle proposed to modify it in various ways, whereas the more rigid version of Bondi and Gold allowed virtually no changes. William McCrea, an early convert to the theory, argued in 1951-53 that continual creation of matter could be understood within the framework of standard general relativity [19]. According to McCrea, matter creation did not really conflict with the law of energy conservation, and the theory also promised a unification of quantum mechanics and cosmology.

As far as observational tests were concerned, the situation remained unsettled for several years. The critics of the steady-state theory believed it could be shot down quickly, but this turned out not to be the case; on the contrary, the theory fared remarkably well. For example, attempts by Allan Sandage and his collaborators to test the theory by comparing its predicted redshift-magnitude relationship with observations failed to yield an unambiguous result. Different world models have different deceleration parameters, and the steady-state value of $q_0 = -1$ distinguished the theory from most relativistic models. By measuring the redshift and magnitude of distant galaxies it would in principle be possible to determine $q_0$, and hence to decide if the steady-state theory was allowed. However, although Sandage concluded that $q_0 > -1$, his data were not good enough to clearly rule out the theory.

The most serious challenge came from the new science of radio astronomy, and in the late 1950s data of radio sources obtained by Martin Ryle's group in Cambridge indicated an incurable disagreement with the prediction of the steady-state theory. A few years later, when the data had stabilized, nearly all radio astronomers agreed that they provided conclusive evidence against the steady-state theory. The supporters of the theory responded by producing alternative explanations of the radio source counts or by suggesting modified steady-state versions designed to cope with the problems. For a short while these responses kept the theory alive, but not much more.

In 1965 Robert Wilson and Arno Penzias unexpectedly detected an isotropic cosmic background of microwaves at wavelength 7.3 cm, which immediately was interpreted as a relic of the hot big bang. In fact, the background radiation had been predicted by Ralph Alpher and Robert Herman as early as 1948, but without attracting any attention [18]. The sensational discovery of 1965 had no natural explanation within the framework of classical steady-state theory. In effect, the cosmic microwave background killed an already dying theory. However, the refutation of the classical steady-state theory, whether in the Hoyle version or the Bondi-Gold version, did not imply that the general idea of an eternally expanding universe with continual creation of matter had to be abandoned.

### 3.  Models from the 1950s and 1960s

During the two decades following World War II a few scientists devised cosmological models which, in a sense, combined features of the relativistic evolution theory with those of the new steady-state theory. The models had in common that they avoided cosmic singularities and a universe temporally closed in either the past or the future. However, this is about all they shared. Being independent models with no convincing observational support, they failed to attract much attention.

In 1952 the British astronomer George McVittie suggested a new world model which drew on results recently obtained by McCrea within the framework of a revised steady-state model [20; 21]. In order to explain the mysterious creation of matter postulated by the steady-state theory, McCrea introduced a zero-point stress in space corresponding to a negative pressure $p = -\rho c^2$, where $\rho$ is the matter density and $c$ the velocity of light. McVittie took over this idea in his model of the "gravitationally steady state" which was, he proved, fully consistent with general relativity. He studied in particular a universe with an open past and an open future (that is, one which has existed for all times), but with a scale factor $R(t)$ that had a non-zero minimum at some time $t = 0$. Throughout the history of the model, from $t = -\infty$ to $t = +\infty$, the pressure was taken to be negative. For $t < 0$ the universe was contracting, with matter being converted into stress, and for $t > 0$ it was expanding, with stress being transformed into matter. It was the latter process that, within the framework of general relativity, would appear to correspond to the continual creation of matter, for which McVittie calculated a rate of the order $10^{-47}$ g/cm$^3$ s.

In spite of having matter creation and no beginning in time, the model did not belong to the steady-state class. The Hubble parameter and the rate of matter creation depended on the cosmic epoch, which meant that it did not satisfy the perfect cosmological principle. McVittie seems not to have considered his model as a candidate for the real universe. Its interest, he wrote, "lies in the fact that model universes of this kind contain the analogue in general relativity of the creation of matter hypothesis of Bondi and Gold" [22, p. 133].

Another kind of model, without matter creation, was suggested by William Bonnor, an orthodox relativist who was equally dissatisfied with the steady-state model and the models of the big-bang type (which he considered to be plainly unscientific). Rather than accepting a primordial singularity, Bonnor wanted an eternal and singularity-free universe consistent with the theory of general relativity. In papers of 1954 and 1957, which primarily dealt with galaxy formation, he suggested that under certain circumstances a contracting Friedmann model could change into an expanding one without passing through a singular state [23; 24]. This would be possible, he thought, if radiation was converted into matter during the contracting phase and a negative pressure was built up near the minimum. Bonnor's favoured candidate was thus an oscillating or cyclical model in which the universe oscillated smoothly, i.e. with no big bangs and no big squeezes (Figure 1). The model had some of the conceptual advantages of the steady-state theory, such as avoiding the question of the creation of the universe, while at the same time keeping on the firm ground of general relativity.

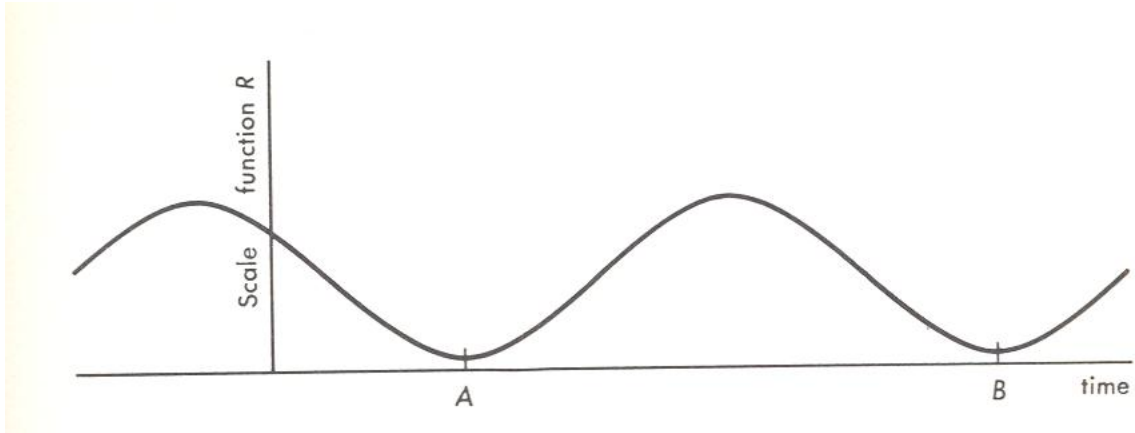

Figure 1. Bonnor's eternally cyclic universe without singular bouncing points. Source: Wm. Bonnor, *The Mystery of the Expanding Universe* (New York: Macmillan, 1964), p. 121.

An idea somewhat related to McVittie's, but not building on standard relativity, was proposed in 1963 by António Gião, a Portuguese mathematical physicist [25]. Based on a generalization of the relativistic field equations, he claimed that none of the cosmological models usually discussed were satisfactory. Rather than introducing ad hoc modifications of the field equations, he suggested to replace the hypothesis of constant mass density by the constancy of the density of proper energy, meaning the sum of proper mass and pressure energies. From this starting point – which contradicts the ordinary perfect cosmological principle – Gião was led to a "generalized steady-state model" in the form of a universe oscillating between a minimum and maximum radius (which he gave as $R_{min} \cong 10^{27}$ cm and $R_{max} \cong 2,8\ R_{min}$). For the period he obtained a value of about $16 \times 10^9$ years. Although matter was created in the contracting phase, it would be compensated by a corresponding destruction in the subsequent expanding phase. In Gião's model the Hubble parameter

*H* pulsated in time, attaining a positive maximum near the middle of the expansion phase and a negative minimum near the middle of the contraction phase. He found that the law of the variation of the world radius was compatible with a nearly constant Hubble parameter during the middle parts of the two phases.

In a work of 1960, further developed in 1961, the Czechoslovakian physicist Jaroslav Pachner proposed that the proper meaning of the cosmological constant was that it expressed either continual creation or annihilation of matter, depending on whether $\Lambda > 0$ or $\Lambda < 0$ [26; 27]. However, he decided to keep to the doctrine *ex nihilo nihil fit* and therefore concluded that $\Lambda = 0$. With this choice he ended up with an oscillating, singularity-free universe that had similarities to the model proposed by Bonnor.

Pachner interpreted the square of the Hubble parameter as proportional to a form of non-gravitating mass density $\rho_H$ which corresponded to the kinetic energy of the cosmic expansion. The total density was given by $\rho = \rho_H + \rho_R$, where $\rho_R$ represented ordinary mass and energy. His pulsating universe was driven by reversible changes between the two forms of energy: during the phase of expansion the "Hubble mass" transformed into rest mass, meaning that new stars and nebulae were formed. When the maximum radius of the universe was reached, all the Hubble mass had been consumed, and during the subsequent contraction stars and nebulae would convert into new Hubble mass, the cycles going on endlessly. Pachner found for the half-period of pulsation $T = 22 \times 10^9$ years and for the time from the present to the state of maximum expansion $19 \times 10^9$ years, meaning that we are in the early expansion phase. The closed world that he considered was only a "cell" in a much larger universe consisting of many spatially separated universes embedded in a higher-dimensional cosmical space. "Since there exists no physical interaction under [*sic*] them, they are incapable of being observed, but this does not signify that they do not exist" [26, p. 673].

Although Pachner had originally dismissed continual creation of matter, in 1965 he suggested a new version of his model in which he incorporated McCrea's idea of a negative, universal stress as the cause of matter creation. He examined a model in which the stress varied as the fourth power of the space curvature and found that, in the present epoch of cosmic evolution, it was practically indistinguishable from the finite-age Friedmann universe [28].

## 4. From *C*-field model to pulsating universe

Contrary to Bondi and Gold, Fred Hoyle insisted in his original formulation of the steady-state theory to keep as closely as possible to the framework of general relativity [16; 18]. This he did by introducing in Einstein's field equations a substitute for the cosmological constant, a symmetric "creation tensor" $C_{\mu\nu}$ the effect of which was to account for the matter creation that was necessary in steady-state cosmology:

$$ R_{\mu\nu} - \frac{1}{2} g_{\mu\nu} R + C_{\mu\nu} = -\kappa T_{\mu\nu} \, , $$

In 1960 he revised the theory by relating the *C*-field to processes of fundamental physics in such a way that the theory appeared in a fully covariant form [29]. This was the first of several revisions that eventually turned the steady-state theory into a version that differed very considerably from its starting point in 1948. For example, whereas the Hubble constant had originally been seen, in conformity with the perfect cosmological principle, as a true constant, in the 1960 formulation Hoyle was ready to let it vary in time.

A few years later the theory was further developed by Hoyle and Jayant Narlikar who now introduced the $C$-field as a scalar quantity associated with a negative energy density. They suggested a mechanism in which matter creation complied with conservation of energy, essentially by having the negative energy of the $C$-field compensate the positive energy of the created particles [30; 31]. The negative energy density acted as a repulsive effect, or an internal pressure, and for this reason singularities would not occur. With the new theory the perfect cosmological principle was definitively abandoned and the universe was pictured as having developed into an exponentially expanding steady state from some initial state. In this asymptotical situation creation and expansion would be in exact balance, given by the same creation rate as in the original steady-state theory, namely $3\rho H$. In the later version of the Hoyle-Narlikar theory [32] it included a gravitational constant decreasing with the cosmic era, $G(t) \sim t^{-1}$, such as Paul Dirac had originally proposed in 1937.

Whereas Hoyle and Narlikar had hitherto followed the standard assumption of a smoothed-out, homogeneous universe, in 1966 they abandoned this assumption, too, and with it the idea of matter being created uniformly through space [33]. As an alternative, they explored the possibility of discrete matter creation around existing concentrations of supermassive bodies, so-called multiplicative creation leading to matter pouring into space. In attempts to calculate the helium content of the universe from non-big bang assumptions, Hoyle developed at the same time an interest in hypothetical supermassive objects that were in some cases acting like miniature oscillating universes. According to Hoyle and Narlikar, as $t \to \pm \infty$, the density would symmetrically approach the steady-state limit $3H^2/4G$, and in between, at $t = 0$, it would attain a minimum given by half this quantity, or the same density as in the original steady-state theory. Although their universe was completely time-symmetric, they argued that an observer would nonetheless experience a particular direction of time.

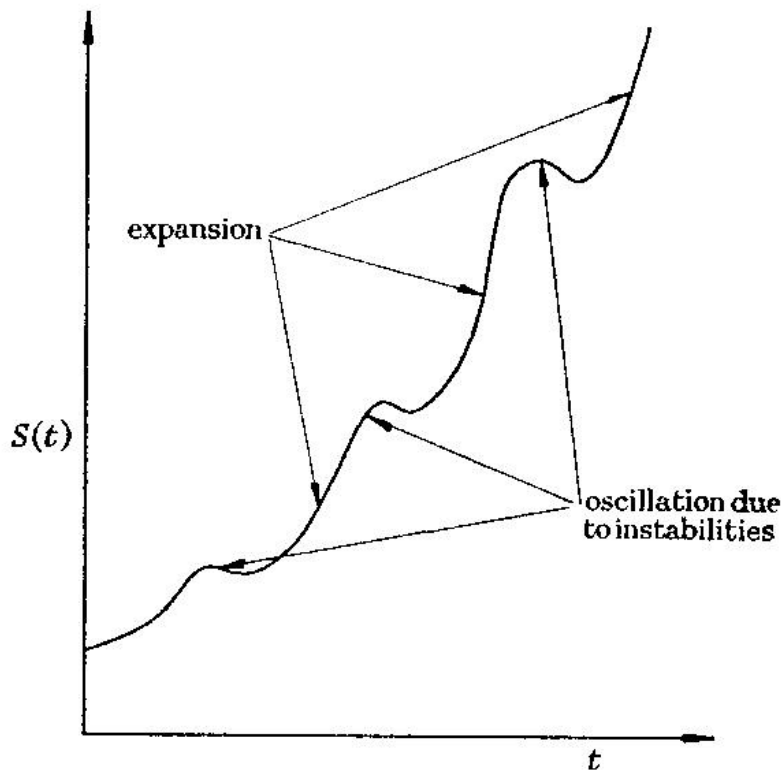

Figure 2.  The Hoyle-Narlikar modified steady state universe [34] with periods of expansion and contraction corresponding to different strengths of the creation process.

In another paper of 1966, Hoyle and Narlikar suggested a model of the universe as consisting of separate bubble universes in which the creation process was temporarily cut off and which would therefore expand much more rapidly than their surroundings [34]. The bubbles would eventually be filled with matter and *C*-fields from the denser outside, but other evacuated regions would be formed at the same time and develop as new bubble universes. If the bubbles developed synchronously, Hoyle and Narlikar found that the universe would follow a series of expansions and contractions; not as in oscillating models, but developing around an exponential steady-state expansion. The possibility of bubble universes led Hoyle into a speculative mood [35, p. 287]:

> The structure we ordinarily think of as "the Universe" is but part of a larger whole. ... Because the fields are generated from the boundary, the particle masses must start there from zero, which is the property that has caused us hitherto to think of the boundary as a singularity. It is natural to ask – If there is a larger Universe "outside", can we learn anything of its structure? What kind of information could we use to come to grips with such a question? … The physical laws themselves – as we determine them experimentally – may have much to tell us about the nature of the larger Universe.

Hoyle and Narlikar expressed great confidence in the pulsating universe built on *C*-field theory, which they considered a strong alternative to the victorious big-bang theory. However, during most of the 1970s and 1980s, when the hot big-bang theory enjoyed nearly universal acceptance, they kept a low profile and did not develop their alternative to any extent.

The general idea of the Hoyle-Narlikar bubble-universe model, that our observable universe is a non-representative part of the universe as a whole, appeared in a different way in a model proposed in 1963 by R. G. Giovanelli. Inspired by "the aesthetic attractiveness of a self-perpetuating universe" Giovanelli was led to abandon the homogeneity postulate and assume that neither the observed expansion nor the mass density of our local region is representative of the entire universe. Neither was the space curvature a constant of the universe as a whole, but merely a description of local phenomena. "It is conceivable that some regions may be expanding, some contracting (according to local observers), and that this state of affairs may reverse from time to time" [36, p. 462]. To account for the large-scale density fluctuations without matter creation, he was led to assume that the cosmological constant was given by $\Lambda = 4\pi G \rho_0$, where $\rho_0$ is the mean density of the universe.

After 1965, any alternative cosmological theory had to account for the cosmic microwave background and its nearly Planckian spectrum. The failure of doing so was a major reason, if not the only reason, why the original steady-state theory was abandoned. Hoyle's answer, which he developed together with Narlikar and Chandra Wickramasinghe in the mid-1970s, was to suggest the existence in interstellar space of thermalizing grains that might convert starlight into long wavelengths of a blackbody-like spectral shape. From this point of view the radiation detected by Penzias and Wilson was not of cosmological origin at all, but merely local starlight dressed into the shape of heat radiation. Initially Hoyle and Wickramasinghe suggested dust grains in the form of graphite needles, but it was eventually realized that slender iron needles or "whiskers," or a combination of the two, might do a better job. At any rate, the dust grain hypothesis was believed to be more natural than the one based on a big bang: "A man who falls asleep on the top of a mountain and who wakes in a fog does not think he is looking at the origin of the Universe. He thinks he is in a fog" [37, p. 810].

By the late 1980s, Hoyle and his few followers had not yet formulated a full alternative to the dominant big-bang theory, but they possessed most of the elements of which the quasi-steady-state model was soon to be constructed. Even though they had abandoned the steady-state theory based on the perfect cosmological principle, they remained strongly opposed to the "dogma" of the big bang. The concerted attack of 1990 that Hoyle, Narlikar, Wickramasinghe, Geoffrey Burbidge, and Halton Arp directed against the big-bang theory in the pages of *Nature* was primarily concerned with observational difficulties [37]. While they hoped to weaken confidence in the big-bang theory, they did not, at this stage, present a viable alternative except that they mentioned the Hoyle-Narlikar theory as a possibility. Three years later the new alternative followed, developed by Hoyle, Burbidge, and Narlikar.

## 5. Quasi-steady-state cosmology

In a series of papers appearing 1993-95 Hoyle, Burbidge, and Narlikar presented a new model of the universe, which they called the quasi-steady-state cosmology or QSSC [38; 39; 40; 41]. In the monograph *A Different Approach to Cosmology*, published in 2000, they reviewed the status of observations and contrasted the standard hot big-bang model with their favourite alternative, arguing that QSSC was in many ways, observationally as well as theoretically, a superior theory of the universe. In their methodological critique of the standard model, Hoyle, Burbidge, and Narlikar repeatedly complained about its many "epicycles," a reference to what they considered to be ad hoc modifications grafted on the model (such as inflation and non-baryonic dark matter). Another standard objection, taken over from the classical steady-state theory, was that the big bang is in principle beyond scientific explanation and can only be accounted for supernaturally.

The universe of QSSC is eternal and creative. Because of the negative pressure provided by the $C$-field, matter is created in conformity with the law of energy conservation. In 1993 Hoyle, Burbidge, and Narlikar described matter creation as taking place in exceedingly violent events of "little big bangs" involving masses as large as $10^{16}$-$10^{17}$ times the mass of the sun, but in their later works they assumed that the creation events were slower and much less violent. "One can say that the universe expands because of the creation of matter," they wrote [41, p. 207]. Although on a very long time scale the size of the observable universe, as given by the scale factor $R$, increases exponentially like in the classical steady-state theory, there is superposed on this expansion oscillations of a smaller time scale. The temporal variation of the scale factor follows an expression such as

$$R(t) = \exp\left(\frac{t}{P}\right)\left[1 + \alpha\cos\left(\frac{2\pi}{Q}\right)\right]$$

where $P \gg Q$ and $\alpha$ is a constant in the range $0 < \alpha < 1$ which signifies the amplitude of the oscillations. The time parameter $Q$ is the cyclical period and $P$ is a measure of the exponential growth of $R$ on a very long time scale. The difference between QSSC and the original steady-state theory is given by the last, oscillatory term. For the two cosmic time scales Hoyle, Burbidge, and Narlikar found in 1994 that $P = 20Q = 8 \times 10^{11}$ years and for the constant $\alpha$, which indicates how close to zero the minimum $R$ can be, they adopted the value 0.75. The cyclical period is thus $Q = 40 \times 10^9$ years, and our epoch is at $t_0 = 0.85Q$ after the last maximum. For this value and the present Hubble parameter $H_0 = 65$ km/s Mpc they derived a present mass density of $\rho_0 \cong 1.79 \times 10^{-29}$ g/cm$^3$ (however, because of the oscillations neither $H$ nor $\rho$ are constant quantities). "The $C$-field provides a negative pressure on which the Universe is able to bounce at the minima of the oscillations ..., and it permits matter to be created to a moderate degree at each such minimum" [40, p. 1017]. Although radiation is produced together with matter, the universe is never radiation dominated, which is one more difference from the big-bang model.

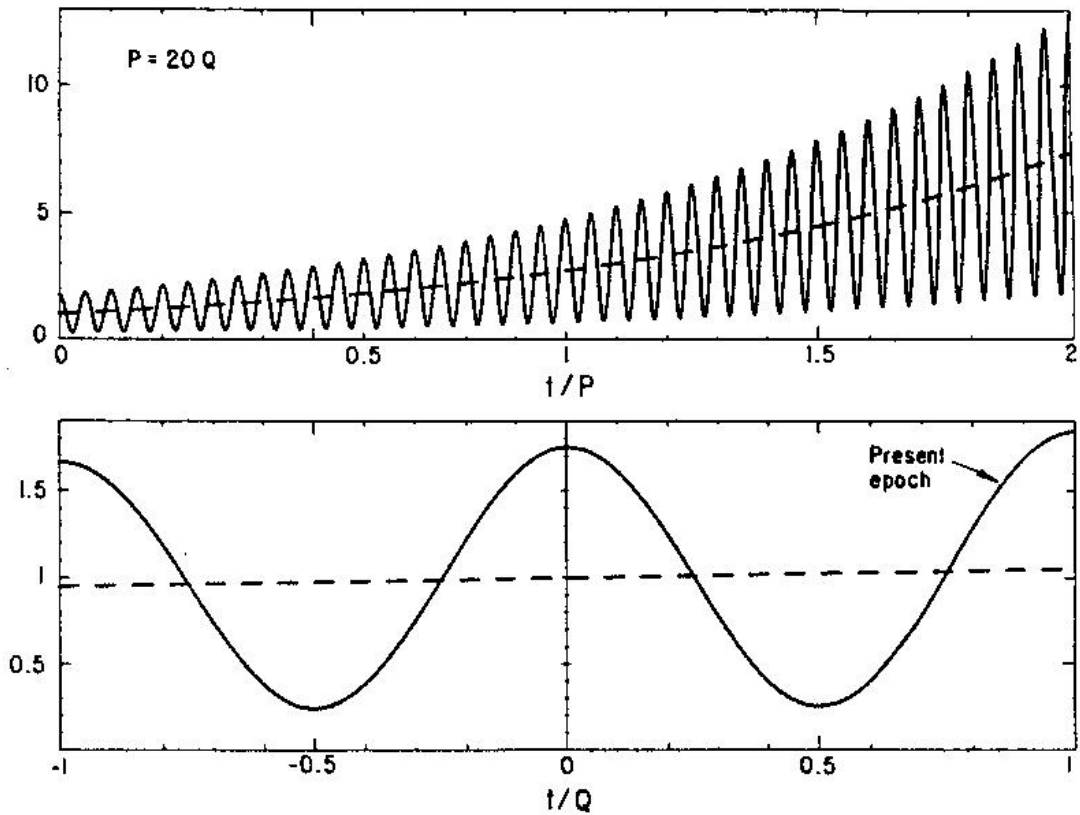

Figure 3. The variation of the scale factor *R* according to the QSSC model with P = 20Q = 8 × 10¹¹ years. *R*, along the vertical axis, is a relative measure of distances in the universe, and the horizontal axis gives the time in terms of P and Q, respectively. The period of oscillations, 4 × 10¹⁰ years, is about three times the age of the big-bang universe.

The energy density of the *C*-field varies as $R^{-4}$, and for this reason it is greatest at the minima. Matter creation is supposed to occur only if the energy-momentum of the *C*-field equals the threshold of momentum of the primordial particle. This happens in epochs when the universe is relatively dense, near the oscillatory minima when the average density is some hundreds of times higher than it is presently. Once matter is created in a "minibang" it is followed by a compensatory negative-energy creation field which is responsible for the explosive character of the creation event and also causes the universe to expand at a high rate. As *R* increases, creation diminishes and the expansion of the universe slows down. As a result of the creation processes, the universe expands and matter creation weakens; the expansion continues until the scale factor reaches the maximum and starts to diminish. The created matter is assumed to appear as highly unstable Planck particles of mass

$$\sqrt{\frac{3c\hbar}{4\pi G}} \cong 10^{-5}\mathrm{g} \cong 10^{19}\mathrm{GeV}$$

where $\hbar = h/2\pi$ is Planck's constant. The Planck particles decay almost instantly to X-bosons from which first quarks and next baryons are formed (of course, antiparticles are formed as well). Each Planck particle produces about $6 \times 10^{18}$ nucleons of ordinary matter within a time of $10^{-24}$ s.

From the theory sketched here Hoyle, Burbidge, and Narlikar derived a number of consequences, one of them being that the present mass density is almost entirely baryonic. Contrary to the situation in the big-bang theory, in QSSC there is no need of large amounts of non-baryonic dark matter. "There is no good astronomical reason to suppose that any of the dark matter is non-baryonic," they concluded; "As far as cosmology is concerned, the existence of non-baryonic matter rests on belief and not on hard evidence" [42, p. 293]. The mechanism of synthesis of the light elements, such as deuterium and helium, largely followed the early-universe scenario in big-bang theory, but not quite. Calculations of the mass fraction of helium-4 gave 0.24, in agreement with observations, and Hoyle, Burbidge, and Narlikar also obtained the correct abundances of deuterium and the isotopes of lithium, beryllium and boron.

QSSC differs from the big-bang theory by predicting that galaxies are formed at all periods and that they contain stars of a wide range of ages. Some stars, born in the previous cycle, are expected to be of ages as large as $40 \times 10^9$ years or more, and thus much older than the age of the universe according to big-bang theory. The theory also predicts a maximum redshift since the last minimum of $z \cong 5$, and although the value may be higher it cannot be very much higher ($z = Hr/c$, where $r$ is the distance to the measured object). Galaxies can be observed at even greater distances, but that does not mean greater redshifts because one will then look to the previous contracting phase. It follows from the theory that we are already seeing into the last oscillatory minimum, where $z \cong 0$. Contrary to the classical steady-state theory, QSSC claimed to have no problem with radio source counts, and it also could reproduce the observed relation between redshifts and angular size. The diameter or angular size of a galaxy (or some other object) varies with the redshift, and the variation follows different expressions according to whether it is based on the steady-state theory or evolutionary cosmologies. This relation as a cosmological test was originally proposed by Hoyle in 1959, and in 1993 Kenneth Kellermann established an empirical relation that he found was consistent with big-bang models of the Einstein-de Sitter type but ruled out the steady-state theory [43]. According to a study of Shyamal Banerjee and Narlikar, a very good fit was provided by QSSC models with low matter density and negative space curvature [44].

What about the cosmic microwave background? The QSSC explanation does not differ much from earlier steady-state explanations, in so far that it appeals to thermalization by means of carbon and metallic whiskers that are synthesized in supernovae. It was of course a problem that there is no direct astronomical evidence for such whiskers, but Hoyle, Burbidge, and Narlikar emphasized that they had been studied for long in the laboratory and that their optical properties were well known. While in the 1993 QSSC theory the microwave background was assumed to have its source in the radiation from the massive creation centres, in the later versions it was explained as a consequence of thermalization of starlight, that is, coming from the fusion of hydrogen into helium. Thermalization is assumed to be carried out in two stages, first by carbon whiskers which convert starlight into infrared light and next by iron whiskers which produce the observed microwave background [45]. Although the spectrum produced by thermalization has almost exactly the form of a blackbody spectrum, it is expected to deviate from the Planck distribution at long wavelengths, about $\lambda > 20$ cm, but the form of the spectral distortions does not follow from the theory.

In 1992 the COBE satellite revealed anisotropies, corresponding to temperature fluctuations, in the cosmic microwave background. The fluctuations were generally interpreted as confirmation of the inflationary big-bang theory, but Hoyle, Burbidge, and Narlikar denied that this was the case. According to QSSC, fluctuations of the size $\Delta T/T \cong 5 \times 10^{-6}$ are expected to arise from slight variations from strict thermalization. They can be reproduced from QSSC assumptions but are not particularly interesting, and in no way do they reveal "the face of God," such as George Smoot said when he presented the COBE data at a press conference in 1992.

The conventional interpretation of the supernovae observations in the late 1990s was that the universe is accelerating, which was generally seen as strong evidence for the existence of a repulsive cosmic force, a "dark energy" due to a positive cosmological constant. However, the observational data can also be explained within the framework of QSSC, such as Hoyle and his collaborators showed in 2000 [46]. With $z_{max} = 5$ and $\alpha = 0.8$ they found a very good fit to the observed redshift-magnitude data for supernovae. As Narlikar, R. Vishwakarma, and Burbidge said two years later, "We feel that the current enthusiasm for a positive cosmological constant may be premature" [47, p. 3]. This is not to say that the constant has no place in QSSC, but it has to be negative and thus contribute to a cosmic deceleration. Reflecting the Machian basis of QSSC, both the cosmological constant and the gravitational constant are determined by the large-scale distribution of matter in the universe, and in such a way that $G > 0$ and $\Lambda < 0$. If $m$ denotes the mass of the created particle and $N$ the number of such particles in the observable universe, the relations are

$$G = \frac{3\hbar c}{4\pi m^2}$$

and

$$\Lambda = -3 \left(\frac{m}{N}\right)^2$$

The observed acceleration does not come from the cosmological constant but from the repulsive force of the negative-energy $C$-field. According to QSSC, the negative cosmological constant, acting as an attractive force, is responsible for the shifts from expansion to contraction that occur when the scale factor is at its maximum value.

The QSSC proposed by Hoyle, Burbidge, and Narlikar in 1993 is still being investigated by Narlikar and a small group of researchers [48]. It was meant to challenge the standard big-bang theory, but only few mainstream cosmologists took up the challenge or showed any interest in the alternative. From a sociological point of view it has clearly been a failure, as indicated by the modest amount of citations to the papers of 1993-95 due to scientists outside the QSSC group. The theory was noticed by the American physicist Peter Phillips who in 1994 proposed a closed steady-state model in which matter was not only created but also annihilated in a different region of the universe [49]. Among the relatively few who responded to the theory was also the American astronomer Edward Wright, who acted as a referee for one of the QSSC papers [40] and was clearly dissatisfied that it was published. In a subsequent paper he claimed not only that the QSSC was disproved by counts of radio sources but also that Hoyle, Burbidge, and Narlikar knew this [50; 51]. Wright has continued to criticize the theory and accused its proponents of scientific dishonesty [52]. Most other mainstream cosmologists have chosen to ignore it, but see Edwin Salpeter's interesting comments in [53].

Supporters of QSSC have, undoubtedly correctly, pointed out that sociological and ideological factors contribute to the lack of interest in this and other alternative theories. As Narlikar and Burbidge complained, "no time has ever been assigned on the Hubble Space Telescope to observers who are thought to fabour the QSSC, and everyone who designs, plans, builds, and observes for the microwave background believes from the beginning that it originated in the big bang" [65, p. 249]. "So much for unbiased observers," they concluded with some bitterness. In a refreshingly self-critical review Narlikar and T. Padmanabhan refer to "the snowball effect arising from the social dynamics of research funding," but they also admit major methodological and technical weaknesses in QSSC. The model "lacks the simplicity and beauty of the original steady state model" and "cannot continue to survive and hope to make an impact based purely on the shortcomings of SC [the standard cosmology]." It needs to be developed to such a level that it can make clear predictions rather than provide post facto explanations for already observed phenomena [54, p. 239 and p. 246].

## 6. Some other eternal-universe models

At the time that QSSC appeared on the cosmological arena, the inflationary big-bang model enjoyed general support and was considered the most promising theory for the very early universe. However, not all cosmologists accepted the theory. In addition to QSSC, several other alternatives were proposed in the 1990s, some of them belonging to the class of eternal universes [66]. One of the alternatives, proposed in 1989 by the Israeli physicists Mark Israelit and Nathan Rosen (the latter of EPR fame), was an attempt to revive the old idea of a cyclical world [55; 56]. Complaining that "discussions of cosmology have become very complicated in recent years, ... [and] the models proposed are highly speculative" [55, p. 627] they suggested a simple model of the early universe that built solely on classical general relativity. The Israelit-Rosen model was spatially closed, cyclical, and avoided singularities. The basic idea was proposed in a paper of 1985, where Rosen introduced the assumption of a limit of matter density $\rho_0$ that will prevent the initial singularity. Referring to the curvature parameter $k$, he summarized his idea as follows: "With the universe having a finite density and scale parameter at this moment [$t = 0$], it is possible to consider also times before $t = 0$. One is led to a model symmetric with respect to the sign of $t$. For the cases $k = 0, -1$, the model first contracts from $\rho = 0$, $R = \infty$ at $t = -\infty$ to $\rho = \rho_0$, $R = R_0$ at $t = 0$ and then expands. For $k = +1$, the model undergoes periodic expansions and contractions from $t = -\infty$ to $t = +\infty$ between $R_0$ and some much larger $R_m$" [57, p. 349].

According to Israelit and Rosen, the universe "began" at $t = 0$ with a hyperdense, tiny "cosmic egg" assumed to have a density of the order of the Planck density $\rho_{Pl} = c^5/\hbar G^2 \cong 10^{93}$ g/cm³. This kind of extraordinary matter ("prematter") is assumed to have a negative pressure, $p = -\rho c^2$, and because of the tension it will result in a large explosion that in some ways corresponds to the inflation phase in the ordinary big-bang theory. Whereas the initial radius of the cosmic egg is about $10^{-34}$ cm, after a time of $10^{-42}$ s it has increased to a value of roughly $10^{-5}$ cm. At about this time the prematter period ends and between $t \cong 10^{-42}$ s and $t \cong 10^{-38}$ s the prematter presumably converts into ordinary particles and radiation. A deceleration sets in and the universe evolves in much the same way as in conventional big-bang theory, including a change from radiation to matter dominance. It continues to grow for the half-period $6 \times 10^{11}$ years, until $R_{\max} \cong 10^{29}$ cm, and then starts to contract until the prematter state has been recreated, which builds up a new tension, etc. The cycles go on endlessly and have always done so (in spite of $t = 0$, there is no beginning of the universe).

Israelit and Rosen offered their cyclical model, based on the hypothesis of a maximally dense state of matter endowed with negative pressure, as an alternative to inflation theory. As they showed, the model was able to take care of the horizon problem and also the flatness problem. Apart from the small segments of time around $t = 0$ (and generally $t = \pm nT$, where $T$ is the period and $n = 0, 1, 2, ...$) it was similar to Friedmann's original cyclical model of 1922, but contrary to this model it included no singularities. One may say that the Israelit-Rosen model was the realization of the kind of non-singular, oscillating universe that Bonnor had considered in the 1950s. The simple model proposed by Israelit and Rosen was developed by a few other researchers who found the model attractive and extended it with a field-theoretical description of the material content of the primeval cosmic egg [58; 59].

A model proposed by Eckhard Rebhan [60] has in common with the Israelit-Rosen model that it starts with a tiny, classical-relativistic micro-universe ($R \cong 10^{-24}$ cm) and thus includes no singularity. However, this models does not exhibit an oscillatory behaviour. Rebhan claims that his model is eternal because the micro-universe is supposed to have always existed, but it is hard to see how one can define time in a changeless proto-universe. From a qualitative point of view Rebhan's "soft bang" model may bring to mind the first big-bang model ever, proposed by Georges

Lemaître in 1931. Lemaître imagined the original universe to consist of a dense and super-radioactive "primeval atom" which at some time exploded. The difference from Rebhan's model is merely that Lemaître's cosmic egg was larger and less dense, of a matter density corresponding to that of an atomic nucleus, roughly $10^{15}$ g/cm³. It was inaccessible to scientific inquiry, devoid of physical properties, and hence non-existent from a physical point of view. It makes no sense to speak of Lemaître's "fireworks universe," or Rebhan's micro-universe, as eternal (nor, really, that it started at "some time"). One should rather say that time originated together with the explosion.

String theory operates with tiny objects (superstrings or branes) of a size close to the Planck length. The strings are described in a 10-dimensional space time with six more space dimensions than those observed. Since they have a non-zero size and cannot collapse to an infinitesimal point, string-based cosmologies provide a way to eliminate the initial singularity. The incompressible string object may perhaps be said to correspond to the cosmic egg in the classical theory of Israelit and Rosen. Among the modern cyclical models the one proposed by Paul Steinhardt and Neil Turok in 2002 has received much attention. This theory, which they speak of as a "new paradigm," was originally inspired by string theory but can be described almost entirely in terms of conventional field theory [61; 67]. Although eternally oscillating between bangs and bounces, this "universe that is made and remade forever" is flat and homogeneous. During a crunch-to-bang transition the ordinary three space dimensions remain finite and so do temperature and density. Although the Steinhardt-Turok cyclical model is entirely novel, it is interesting to notice that it – like some of the earlier cyclical models – was motivated by dissatisfaction with inflation big-bang theory.

## 8. Concluding remarks

Among the numerous conceptions of the universe that have been proposed through history, one may distinguish between two large groups, depending on whether the age is taken to be finite or infinite. In most of the post-World War II period, finite-age models have been the most popular, and since the 1960s they have completely dominated cosmological thought in the form of the hot big-bang paradigm. However, although ideas of the eternal universe have for a long time been marginalized, they have lived on in various versions. Most of these versions are dynamic, in the sense that they accept the expansion of the universe, but there are also models that depict the universe as static. In 1988 the American historian of physics Gerald Holton called attention to the historical invariance of the themes underlying the pair of contrasting cosmologies in the shape of, respectively, eternal-cyclic world pictures and creation world pictures. Holton noted the apparent final triumph of the big-bang creation cosmogony over the rival view based on eternal existence, but he also prophesied that "this thema will come in again through the back door" [62, p. 46]. Indeed, this is what has happened during the last 15 years or so.

Although everyone agree that ultimately there is only one valid way to evaluate cosmological theories – according to how well they fare in observational tests – it is obvious that such tests are not alone in deciding the fate or popularity of a theory. There are other factors at play, some of them of a metaphysical and emotional kind. Scientists may feel a certain theory to be "attractive" even though it does not fare well with regard to observations. To illustrate this point, let me end by quoting from the lecture that Andrei Sakharov, the Russian physicist and political dissident, delivered (*in absentia*) when he was awarded the Nobel peace prize in 1975. Sakharov, who in the late 1960s did very important work in theoretical early-universe cosmology, said as follows [63, p. 32]:

> I support the cosmological hypothesis that states that the development of the universe is repeated in its basic characteristics an infinite number of times. ... This weltanschauung cannot in the least devalue our sacred inspirations in this world, into which, like a gleam in the darkness, we have appeared in the darkness for an instant from the black nothingness of the ever-conscious matter, in order to make good the demands of reason and create a life worthy of ourselves and of the goal we only dimly perceive.

## Bibliography

1. Rey, A. (1927): *La retour éternel et la philosophie de la physique*. Flammarion, Paris.
2. Jaki, S. (1974): *Science and creation: From eternal cycles to an oscillating universe*. Scottish Academic Press, Edinburgh.
3. O'Brien, D. (1969): *Empedocles' cosmic cycle. A reconstruction from the fragments and secondary sources*. Cambridge University Press, Cambridge.
4. Sambursky, S. (1963): *The physical world of the Greeks*. Routledge and Kegan Paul, London.
5. Kragh, H. (2004): *Matter and spirit in the universe*. Imperial College Press, London.
6. Köhler, O. (1895): *Weltschöpfung und Weltuntergang*. Diess Verlag, Stuttgart.
7. Kragh, H. (1995): Cosmology between the wars: The Nernst-MacMillan alternative. *J. Hist. Astron.* 26, 93-115.
8. MacMillan, W. (1925): Some mathematical aspects of cosmology. *Science* 62, 63-72, 96-99, 121-127.
9. Friedmann, A. (2000): *Die Welt als Raum und Zeit*. Ed. by G. Singer. Harri Deutsch, Frankfurt am Main.
10. Tolman, R. (1934): *Relativity, thermodynamics, and cosmology*. Oxford University Press, Oxford.
11. Lemaître, G. (1933): L'Univers en expansion. *Ann. Soc. Sci. Bruxelles* 53, 51-85 (English trans. in *Gen. Rel. and Grav.* 29, 1997, 641-680).
12. Eddington, A. (1928): *The Nature of the physical world*. Cambridge University Press, Cambridge.

13. Einstein, A. (1931): Zum kosmologischen Problem der allgemeinen Relativitätstheorie. *Sitzungsber. Preuss. Akad. Wiss.*, 142-152.

14. Spencer Jones, H. (1934): *General astronomy*. Cambridge University Press, Cambridge.

15. De Sitter, W. (1931): *Nature* 138, 708.

16. Hoyle, F. (1948): A new model for the expanding universe. *Month. Not. Roy. Astron. Soc.* 108, 372-382.

17. Bondi, H., Gold, T. (1948): The steady-state theory of the expanding universe. *Month. Not. Roy. Astron. Soc.* 108, 252-270.

18. Kragh, H. (1996): *Cosmology and controversy: The historical development of two theories of the universe*. Princeton University Press, Princeton.

19. McCrea, W. (1951): Relativity theory and the creation of matter. *Proc. Roy. Soc. A* 206, 562-575.

20. McVittie, G. (1952): A model universe admitting the interchangeability of stress and matter. *Proc. Roy. Soc. A* 211, 295-301.

21. Kragh, H. (1999): Steady-state cosmology and general relativity: Reconciliation or conflict? In: Goenner, H. et al. (eds.): *The expanding worlds of general relativity*. Birkhäuser Verlag: Basel, 377-402.

22. McVittie, G. (1953): The age of the universe in the cosmology of general relativity. *Astron. J.* 58, 129-134.

23. Bonnor, W. (1954): The stability of cosmological models. *Zs. Astrophys.* 35, 10-20.

24. Bonnor, W. (1957): La formation des nébuleuses en cosmologie relativiste. *Ann. Inst. Poincaré* 15, 158-172.

25. Gião, A. (1963): On the theory of the cosmological models with special reference to a generalized steady-state model. In: Gião, A. (ed.): *Cosmological models*. Instituto Gulbenkian de Ciência: Lisbon, 5-100.

26. Pachner, J. (1960): Dynamics of the universe. *Acta Phys. Pol.* 19, 662-673.

27. Pachner, J. (1961): Zur relativistischen Kosmologie. *Ann. Phys.* 8, 60-75.

28. Pachner, J. (1965): An oscillating isotropic universe without singularity. *Month. Not. Roy. Astron. Soc.* 131, 173-176.

29. Hoyle, F. (1960): A covariant formulation of the law of creation of matter. *Month. Not. Roy. Astron. Soc.* 120, 256-262.

30. Hoyle, F., Narlikar, J. (1962): Mach's principle and the creation of matter. *Proc. Roy. Soc. A* 270, 334-341.

31. Hoyle, F., Narlikar, J. (1964): The C-field as a direct particle field. *Proc. Roy. Soc. A* 282, 178-183.

32. Narlikar, J. (1983): *Introduction to cosmology*. Cambridge University Press, Cambridge.

33. Hoyle, F., Narlikar, J. (1966): On the effects of the non-conservation of baryons in cosmology. *Proc. Roy. Soc. A* 290, 143-160.

34. . Hoyle, F., Narlikar, J. (1966): A radical departure from the 'steady-state' concept in cosmology. *Proc. Roy. Soc. A* 290, 162-176.

35. Hoyle, F. (1973): The origin of the universe. *Quart. J. Roy. Astron. Soc.* 14, 278-287.

36. Giovanelli, R. (1963-64): A fluctuation theory of cosmology. *Month. Not. Roy. Astron. Soc.* 127, 461-469.

37. Arp, H. et al. (1990): The extragalactic universe: An alternative view. *Nature* 346, 807-812.

38. Hoyle, F., Burbidge, G., Narlikar, J. (1993): A quasi-steady state cosmological model with creation of matter. *Astrophys. J.* 410, 437-457.

39. Hoyle, F., Burbidge, G., Narlikar, J. (1994): Further astrophysical quantities expected in a quasi-steady state universe. *Astron. Astrophys.* 289, 729-739.

40. Hoyle, F., Burbidge, G., Narlikar, J. (1994): Astrophysical deductions from the quasi-steady state cosmology. *Month. Not. Roy. Astron. Soc.* 267, 1007-1019.

41. Hoyle, F., Burbidge, G., Narlikar, J. (1995): The basic theory underlying the quasi-steady-state cosmology. *Proc. Roy. Soc. A* 448, 191-212.

42. Hoyle, F., Burbidge, G., Narlikar, J. (2000): *A different approach to cosmology*. Cambridge University Press, Cambridge.

43. Kellermann, K. (1993): The cosmological deceleration parameter estimated from the angular-size/redshift relation for compact radio sources. *Nature* 361, 134-136.

44. Banerjee, S., Narlikar, J. (1999): The quasi-steady-state cosmology: A study of angular size against redshift. *Month. Not. Roy. Astron. Soc.* 307, 73-78.

45. Wickramasinghe, C. (2006): Evidence for iron whiskers in the universe. In: Pecker, J.-C., Narlikar, J. (eds.): *Current Issues in Cosmology*. Cambridge University Press: Cambridge, 152-163.

46. Banerjee, S. et al. (2000): Possible interpretations of the magnitude-redshift relation for supernovae of type Ia. *Astrophys. J.* 119, 2583-2588.

47. Narlikar, J., Vishwakarma, R., Burbidge, G. (2002): Interpretations of the accelerating universe. Astro-ph/0205064.

48. Narlikar, J. (2006): The quasi-steady-state cosmology. In: Pecker, J.-C., Narlikar, J. (eds.): *Current Issues in Cosmology*. Cambridge University Press: Cambridge, 139-151.

49. Phillips, P. (1994): Solutions of the field equations for a steady-state cosmology in a closed space. *Month. Not. Roy. Astron. Soc.* 269, 771-778.

50. Wright, E. (1994): Comments on the quasi-steady-state cosmology. Astro-ph/9410070.

51. Hoyle, F., Burbidge, G., Narlikar, J. (1994). Note on a comment by Edward L. Wright. Astro-ph/9412045.

52. Wright, E. (2006): www.astro.ucla.edu/~wright/stdystat.htm

53. Salpeter, E. (2005). Fallacies in astronomy and medicine. *Rep. Progr. Phys.* 68, 2747-2772.

54. Narlikar, J., Padmanabhan, T. (2001): Standard cosmology and alternatives: A critical appraisal. *Ann. Rev. Astron. Astrophys.* 39, 211-248.

55. Israelit, M., Rosen, N. (1989): A singularity-free cosmological model in general relativity. *Astrophys. J.* 342, 627-634.

56. Israelit, M., Rosen, N. (1991): Perturbations in a singularity-free cosmological model. *Astrophys. J.* 375, 463-472.

57. Nathan, N. (1985): General relativity cosmological models without the big bang. *Astrophys. J.* 297, 347-349.

58. Starkovich, S., Cooperstock, F. (1992). A cosmological field theory. *Astrophys. J.* 398, 1-11.

59. Bayin, S. et al. (1994): A singularity-free cosmological model with a conformally coupled scalar field. *Astrophys. J.* 428, 439-446.

60. Rebhan, E. (2000): 'Soft bang' instead of 'big bang': Model of an inflationary universe without singularities and with eternal physical past time. *Astron. Astrophys.* 353, 1-9.

61. Steinhardt, P., Turok, N. (2002): The cyclic universe: An informal introduction. Astro-ph/020447.

62. Holton, G. (1988): *Thematic origins of scientific thought*. Harvard University Press, Cambridge, Mass.

63. Drell, S., Okun, L. (1990): Andrei Dmitrievich Sakharov. *Phys. Today* 43: 8, 26-36.

64. Kragh, H. (2012): Walther Nernst: Grandfather of dark energy? *Astron. & Geophys*. 53, (forthcoming).

65. Narlikar, J., Burbidge, G. (2008): *Facts and speculations in cosmology*. Cambridge University Press, Cambridge.

66. Kragh, H. (2009): Continual fascination: The oscillating universe in modern cosmology. *Science in Context* 22, 587-612.

67. Steinhardt, P., Turok, N. (2007). *Endless universe: Beyond the big bang*. Doubleday, New York.